\documentstyle[12pt,psfig]{article}

\def\theequation{\arabic{section}.\arabic{equation}}

\renewcommand{\baselinestretch}{1.1}

\begin{document}

{\sc\noindent Research Institute for Theoretical Physics\hfill
University of Helsinki\\Preprint series in Theoretical Physics,\hfill
HU-TFT-96-43, \hfill 25 November 1996}

\vspace*{0.5cm}

\begin{center}
{\Large\bf States prepared by decay}

\vspace*{0.5cm}

{\large Stig Stenholm and Asta Paloviita}

\bigskip

{\it Research Institute for Theoretical Physics\\ and the Academy of Finland\\
P.O.Box 9, FIN-00014 University of Helsinki, Finland} 

\bigskip

{\bf Abstract}

\end{center}

\begin{quote}
We consider the time evolution of a discrete state embedded in a continuum.
Results from scattering theory can be utilized to solve the initial value
problem and discuss the system as a model of wave packet preparation.
Extensive use is made of the analytic properties of the propagators, and
simple model systems are evaluated to illustrate the argument. We verify
the exponential appearence of the continuum state and its propagation as a
localized wave packet.
\end{quote}

\begin{quote}
{\it This article will appear in a Special Issue of Journal of Modern
Optics in 1997}
\end{quote}

\section{Introduction}

Quantum mechanics knows many examples of discrete states coupled to a
continuum. Much work has been devoted to these problems in the past. In
this paper we are going to review these works briefly, but reconsider them
as methods to prepare time dependent states. All tunnelling problems are of
this type, but in addition we have similar phenomena occurring when a
photon excitation is followed by leakage into continuous outgoing states. 

A simple physical example is provided by autoionization, where an excited
state is coupled to atomic ionization according to 
\begin{equation}
   A+\hbar \omega \rightarrow A^{*}\rightarrow A^{+}+e^-. \label{e1}
\end{equation}
In molecules, the analogous process is the predissociation reaction
\begin{equation}
   AB+\hbar \omega \rightarrow AB^{*}\rightarrow A+B, \label{e2}
\end{equation}
and in chemical reactions the transition state may give a similar behaviour
\begin{equation}
   AB+C+\hbar \omega \rightarrow ABC^{*}\rightarrow A+BC. \label{e3}
\end{equation}
In a nucleus, $ABC^{*}$ would be a compound nucleus or a doorway state. In
particle physics, the particle production processes provide additional
examples.

When the final states form a continuous spectrum, the decay of the
initially prepared state is often described by an exponential. This was
first derived from quantum theory by Landau~\cite{r1}, but the procedure is
usually referred to by the names Weisskopf and Wigner~\cite{r2}. These
features were found to appear exactly in the treatment of the Lee 
model~\cite{r3}, where, however, the primary interest was devoted to the
renormalizability of the theory. The appearance of decay as a consequence
of the analytic properties of the propagators became a central issue for
the theoretical discussion after Peierls had pointed out that the functions
needed to be continued to the second Riemann sheet~\cite{r4}. These
discussions can be found in Refs.~\cite{r5} and~\cite{r6}, and the results
are summarized in the context of scattering theory in the text book~\cite{r7}. 

When several singularities are close to each other or multiple poles occur,
there appears non-exponential time dependence as discussed by Mower~\cite{r8}; 
the corresponding problem for a generalized Lee-model is
treated in Ref.~\cite{r9}. In molecules the high density of states gives
rise to a similar physical situation through nonradiative transfer of
excitation; for a review of this theory see Ref.~\cite{r10}. Many more
works from this early period could be cited, but these may suffice for the
present. 

The works above are mainly formulated as scattering problems and, except
for the evaluation of probability decay, they tend not to look into the
detailed time evolution of the processes. The use of well controlled laser
pulses in the femtosecond time range has recently made it possible to
follow microscopic processes in space and time~\cite{ra1}. Thus one may
resolve the Schr\"odinger propagation following a fast preparation of an
initial state. It is then interesting to reconsider the theories above from
the point of view of state preparation. If we excite a resolved initial
state coupled to a continuum, which kind of wave packet can we prepare into
the continuum? How is this state emerging and how is it propagating? 

This can be seen as a continuation of our earlier work on molecular
dynamics~\cite{r11} and on electrons in semiconductor structures~\cite{r12}. 
Here we consider explicitly the time dependence following an
initial preparation. Our state functions can thus stay normalized at all
times in contrast to the situation in a scattering description. Thus we can
avoid paradoxes of exponentially growing amplitudes such as discussed by
Peierls in Ref.~\cite{r13}.

The preparation of well localized wave packets is an interesting and
challenging problem in atomic physics. Various excitation processes have
produced electronic wave packets on bound Rydberg manifolds; see e.g. 
Refs.~\cite{rr0} and~\cite{rr1}, but it is not clear if these can be launched
into freely propagating states. Molecular dissociation, following coherent
excitation from the ground state~\cite{r11}, is expected to prepare
localized wave packets in the reaction coordinate. In simple cases, this
can be seen as propagation in the laboratory too. To prepare electronic
wave packets in semiconductors poses huge experimental difficulties~\cite{rr2};
to retain their coherence seems almost impossible. Excitonic
wave packets dephase more slowly, but their theoretical description is even
more difficult.

In this paper we adopt the presumption that an experimental procedure
exists to prepare a pure isolated state. The interaction with the continuum
is set in, and we can follow both the growth of the continuum states and
the decay of the initial one. This makes it possible to consider the
process as a state preparation. We can describe the disappearance of the
initial state and the growth and propagation of the state prepared in the
continuum. 

In order to achieve our goal, we postulate the validity of a simple model
Hamiltonian. It is of the Lee model type and can hence be solved exactly to
give formal expressions for all quantities of interest. In order to treat
the initial value problem we utilize the Laplace transform, but in order to
retain agreement with the conventional Fourier transform treatments, we
denote the Laplace transform variable by $-i\omega$; the resulting
transform is named the Laplace-Fourier transform
(${\cal LF}$-transform). 

We present the model in Sec.~2 and its formal solution in Sec.~3, where also
the general behaviour of the solution is discussed. In Sec.~4 we present
some simple models which illustrate the features of the general discussion. 

In Sec.~4.1 we assume that the continuum spectral density can be described
by a simple complex pole, the generalization to several poles is
straightforward. This approximation has acquired renewed interest after
Garraway~\cite{r14} has showed that the multiple pole approximation can be
put into a Lindblad form amenable to a Monte Carlo simulation in terms of
hypothetical pseudomodes, which describe the non-Markovian effects.
However, he finds that consistency may require that the pseudomodes are
coupled in their Hamiltonian.

Section~4.2 presents a simple model where the role of analytic continuation
can be elucidated and the exact behaviour of the wave packet in the
continuum can be calculated. The role of the process as a preparation is
clearly seen in this model. In Sec.~4.3 we try to construct a physical 
system having the properties of our earlier models. 
This is selected from the field of molecular 
physics~\cite{r11} and tests how weak coupling to a molecular continuum can be
utilized to prepare a wave packet. We find that this is, indeed, possible,
but the system has got many complementary features which we will discuss in
a forthcoming paper. 

Section~5 presents a brief discussion of our results 
and the conclusions. 

\section{The model calculation}\setcounter{equation}{0}

We are considering a model with one single state embedded in a continuum.
This is described by the Hamiltonian
\begin{equation}
\begin{array}{lll}
H & = & H_0+V \\
& & \\
H_0 & = & \omega_0\mid 0\rangle \langle 0\mid +\int d\epsilon \mid
\epsilon \rangle \,\epsilon \,\langle \epsilon \mid \\ & & \\
V & = & \int d\epsilon \;\left( V_\epsilon \mid \epsilon \rangle \langle
0\mid +V_\epsilon ^{*}\mid 0\rangle \langle \epsilon \mid \right)
\end{array}
\label{a1}
\end{equation}
in an obvious notation. Observe that this is the form of a general
tunnelling Hamiltonian. The initial state is supposed to be the isolated
state
\begin{equation}
\mid \psi (t=0)\rangle =\mid 0\rangle . \label{a2} \end{equation}
Because we are going to consider time evolution, we introduce the
Laplace-Fourier transform in the form
\begin{equation}
\widetilde{\psi }(\omega )=-i\int\limits_0^\infty dt\, e^{i\omega t}\psi
(t)\,\equiv {\cal LF}(\psi ). \label{a3} 
\end{equation}
This gives us analytic functions in the conventional half-plane ${\rm Im}(\omega
)\geq 0$. We have the usual properties
\begin{equation}
\begin{array}{lll}
{\cal LF}(\dot \psi ) & = & -i\left( \omega \widetilde{\psi }(\omega
)-\psi (t=0)\right) \\
& & \\
\lim\limits_{\omega \rightarrow \infty }\left( \omega \widetilde{\psi }%
(\omega )\right) & = & \psi (t=0)\equiv \psi_0 \\ & & \\
\lim\limits_{\omega \rightarrow 0}\left( \omega \widetilde{\psi }(\omega
)\right) & = & \psi (t=\infty ).
\end{array}\label{a4}
\end{equation}
The inversion is achieved by
\begin{equation}
\psi (t)=\frac i{2\pi }\int\limits_{-\infty +ia}^{+\infty +ia}e^{-i\omega
t}\;\widetilde{\psi }(\omega )\,dt, \label{a5} 
\end{equation}
where $a>0$. With these definitions the solution of the Schr\"odinger
equations is given by
\begin{equation}
   \widetilde{\psi }(\omega )=G(\omega )\psi_0, \label{a6} 
\end{equation}
where the propagator is
\begin{equation}
   G(\omega )=\frac 1{\omega -H}. \label{a7} 
\end{equation}
It is easy to verify the consistency of these relations. 

We are now going to partition the problem above in order to separate the
time evolution of the single state and that of the continuum. Thus we
introduce the projectors
\begin{equation}
\begin{array}{lll}
   P & = & \mid 0\rangle \langle 0\mid \\
   & & \\
   Q & = & 1-P=\int d\epsilon \mid \epsilon \rangle \,\langle \epsilon \mid .
\end{array}\label{a8}
\end{equation}
Both projectors commute with the Hamiltonian $H_0.$ 

The results that we need were derived long ago in scattering theory, but
for easy reference we summarize them in the Appendix. We introduce the
definition
\begin{equation}
O^{XY}\equiv XOY, \label{a9}
\end{equation}
where $O$ is any operator and $X,Y=P$ or $Q;$ the operators $O^{XX}$ are
denoted by $O^X$ simply.

We introduce a connected operator $\Gamma $ which has the single state pole
removed from the intermediate states by writing \begin{equation}
\Gamma =V+V\,G_0^Q\,\Gamma , \label{a10} \end{equation}
where the unperturbed propagator is
\begin{equation}
G_0(\omega )=\frac 1{\omega -H_0}. \label{a11} \end{equation}
In terms of this operator, we can now write the exact solution for all the
propagators needed in the form
\begin{eqnarray}
  G^P&=&\left( (G_0^P)^{-1}-\Gamma ^P\right) ^{-1} \label{a12a}\\
  G^{QP}&=&G_0^QV^{QP}G^P \label{a12b}\\
  G^Q&=&G_0^Q+G_0^Q\left( \Gamma ^Q+\Gamma ^{QP}G^P\Gamma ^{PQ}\right) G_0^Q.
\label{a12c}
\end{eqnarray}
Thus we can obtain all desired partitioned propagators if we can solve the
integral equation (\ref{a10}). In the present case, this can be solved
trivially if we notice that
\begin{equation}
  V^P=V^Q=0. \label{a13}
\end{equation}
From this it follows that
\begin{eqnarray}
  \begin{array}{lll}
    \Gamma ^{QP} & = & V^{QP} \\
    & & \\
    \Gamma ^P & = & V^{PQ}G_0^QV^{QP}.
  \end{array}\label{a14}
\end{eqnarray}
This gives the solutions
\begin{eqnarray}
\langle 0\mid G\mid 0\rangle &=&\frac 1{\omega -\omega_0-\Sigma (\omega )}
\label{a15a}\\
\langle \epsilon \mid G\mid 0\rangle &=& \left( \frac 1{\omega -\epsilon
}\right) V_\epsilon \left( \frac 1{\omega -\omega_0-\Sigma (\omega
)}\right) \label{a15b}\\
\langle \epsilon \mid G\mid \epsilon ^{^{\prime }}\rangle &=&\frac{\delta
(\epsilon -\epsilon ^{^{\prime }})}{\omega -\epsilon }+\left( \frac{%
V_\epsilon \,}{\omega -\epsilon }\right) \left( \frac{V_{\epsilon ^{\prime
}}^{*}}{\omega -\epsilon ^{^{\prime }}}\right) \left( \frac 1{\omega
-\omega_0-\Sigma (\omega )}\right) . \label{a15c} 
\end{eqnarray}
The self-energy function is given by
\begin{equation}
  \Sigma (\omega )=\langle 0\mid \Gamma \mid 0\rangle =\int d\epsilon \left(
  \frac{D(\epsilon )}{\omega -\epsilon }\right) , \label{a16} 
\end{equation}
where we have introduced the spectral density \begin{equation}
D(\epsilon )=\mid V_\epsilon \mid ^2\geq 0. \label{a19} \end{equation}
This notation is preferable because we can then include a density-of-states
function in the spectral density $D(\omega )$ when needed. 

With the initial condition (\ref{a2}) we find the amplitude of the isolated
state to be
\begin{equation}
A(t)\equiv \langle 0\mid e^{-iHt}\mid 0\rangle =\frac i{2\pi
}\int\limits_{-\infty +ia}^{+\infty +ia}e^{-i\omega t}\;\left( \frac
1{\omega -\omega_0-\Sigma (\omega )}\right) \,dt. \label{a17}
\end{equation}
Because the integrated function is analytic in the upper half plane, the
time evolution is determined by the singularities of the integrand on the
real axis and below. We are thus looking for solutions of the equation
\begin{equation}
h(\omega )\equiv \omega -\omega_0-\Sigma (\omega )=0. \label{a18}
\end{equation}
As can be seen from Eq.~(\ref{a16}) the function $\Sigma (\omega )$ has got
a branch cut along the real axis. Its weight is given by $D(\omega ).$ 

\section{General properties of the solution} \setcounter{equation}{0}

From the definition (\ref{a16}) it follows directly that 
\begin{equation}
{\rm Im}\,\Sigma (\omega \pm i\eta )=\mp \pi D(\omega ); \label{a20}
\end{equation} it is assumed that $\eta \rightarrow 0.$ Consequently, the
function has a branch cut at every value of $\omega $ such that $D(\omega )$
differs from zero.

All physical systems have energies bounded from below, thus all integrals
over the variable $\epsilon $ must start at a finite value $\mu .$ If the
energy of the isolated state lies below this spectral cut-off
\begin{equation}
  \omega_0<\mu , \label{c1}
\end{equation}
the continuum only serves to renormalize the state. Then we may write
\begin{equation}
\Sigma (\omega)=\int d\epsilon \;\frac{D(\epsilon )}{\omega 
-\omega_0-(\epsilon -\omega_0)}=\int d\epsilon \;\frac{D(\epsilon )}
{(\omega_0-\epsilon )}-\int d\epsilon \;\frac{D(\epsilon )}{(\omega_0-\epsilon
)^2} \;\left( \omega -\omega_0\right) + ... \label{c2} 
\end{equation}
This gives the propagator (\ref{a15a}) the form 
\begin{equation}
\langle 0\mid G\mid 0\rangle =\frac{Z}{\omega -\tilde{\omega}_0},
\label{c3}
\end{equation}
where the state renormalization constant is given by 
\begin{equation}
  Z=\left( 1+\int d\epsilon \;\frac{D(\epsilon )}{(\omega_0-\epsilon )^2}
  \right) ^{-1} \label{c4}
\end{equation}
and the renormalized energy is
\begin{equation}
\tilde{\omega}_0=\omega_0+Z\int d\epsilon \;\frac{D(\epsilon )}{
(\omega_0-\epsilon)}<\omega_0. \label{c5} 
\end{equation}
These results are valid to order $\left( \frac{\omega_0-\tilde{\omega}_0 }
{\mu -\omega_0}\right)$, and we still have $\tilde{\omega}_0<\mu$. 
The only effect of the continuum is to push the isolated level
away and decrease the overlap between the initial bare state and the
renormalized state to $Z<1$. This state does not decay but continues
oscillating with the frequency $\tilde{\omega}_0$ forever. The
reminder of the initial state is lost into the continuum as the coupling is
switched on. 

When the isolated state is embedded in the continuum, $\omega_0>\mu$, it
can decay. In the equation (\ref{a17}) we require the zeros of the function
$ h(\omega)$ to be in the lower half plane. Let us see if such
singularities exist by setting
\begin{equation}
\omega =\omega ^{\prime}-i\omega ^{\prime \prime } \label{a21}
\end{equation}
into $h(\omega)$. We find
\begin{equation}
\omega ^{\prime }-i\omega ^{\prime \prime }-\omega_0-\int d\epsilon
\;\frac{ D(\epsilon )}{\omega ^{\prime }-i\omega ^{\prime \prime
}-\epsilon }=0. \label{a22}
\end{equation}
Separating the real and imaginary parts of this equation should fix the
oscillational frequency $\omega ^{\prime }$ and the damping $\omega
^{\prime \prime }.$ We write down the imaginary part as \begin{equation}
\omega ^{\prime \prime }\left( 1+\int d\epsilon \;\frac{D(\epsilon
)}{\left( \omega ^{\prime }-\epsilon \right) ^2+\left( \omega ^{\prime
\prime }\right) ^2}\right) =0. \label{a23}
\end{equation}
This equation obviously lacks solutions $\omega ^{\prime \prime }\neq 0$
and does not correspond to a physical result. This is obtained when we
remember that the physical values of $\omega $ approach the real axis from
the upper half plane, and we should thus look for zeros when the function
$\Sigma (\omega )$ is continued analytically into the lower half plane.
This procedure will provide the singularities giving a contribution to the
integral (\ref{a17}). We consequently need to continue $\Sigma (\omega )$
across the real axis as discussed by Peierls~\cite{r14}. 

The simplest manner to do this is to push the integration contour in
Eq.~(\ref {a17}) down below the real axis. Any isolated pole encountered in
this manner is included as a single contribution to the time evolution in
the manner shown in Fig.~1. For one single pole this works well, and the
result is simple exponential decay as in the Weisskopf-Wigner approach.
However, with several singularities only well isolated poles can be treated
this way. When the poles are close to each other, this method cannot be
justified, because its validity is based on a series expansion like that in
Eq.~(\ref{c2}) around each singularity; the radius of convergence can only be
extended to the nearest singularity, and hence close lying poles 
interfere~\cite{r8}. Another limitation comes from the lower cut-off 
in the spectral
weight $ D(\omega)$. Because $D(\omega)=0$ for $\omega <\mu $, this end
point must be a singularity of the function. Thus there is a branch cut
starting here, which can be moved around but must be pinned down at $\omega
=\mu$. Poles too close to this cut-off cannot have a large radius of
convergence and hence the branch cut cannot be neglected in their
treatment. In particle theory these cuts derive from particle creation
thresholds. 

\begin{figure}
\vspace*{-0.5cm}
\centerline{\psfig{width=2.4in,file=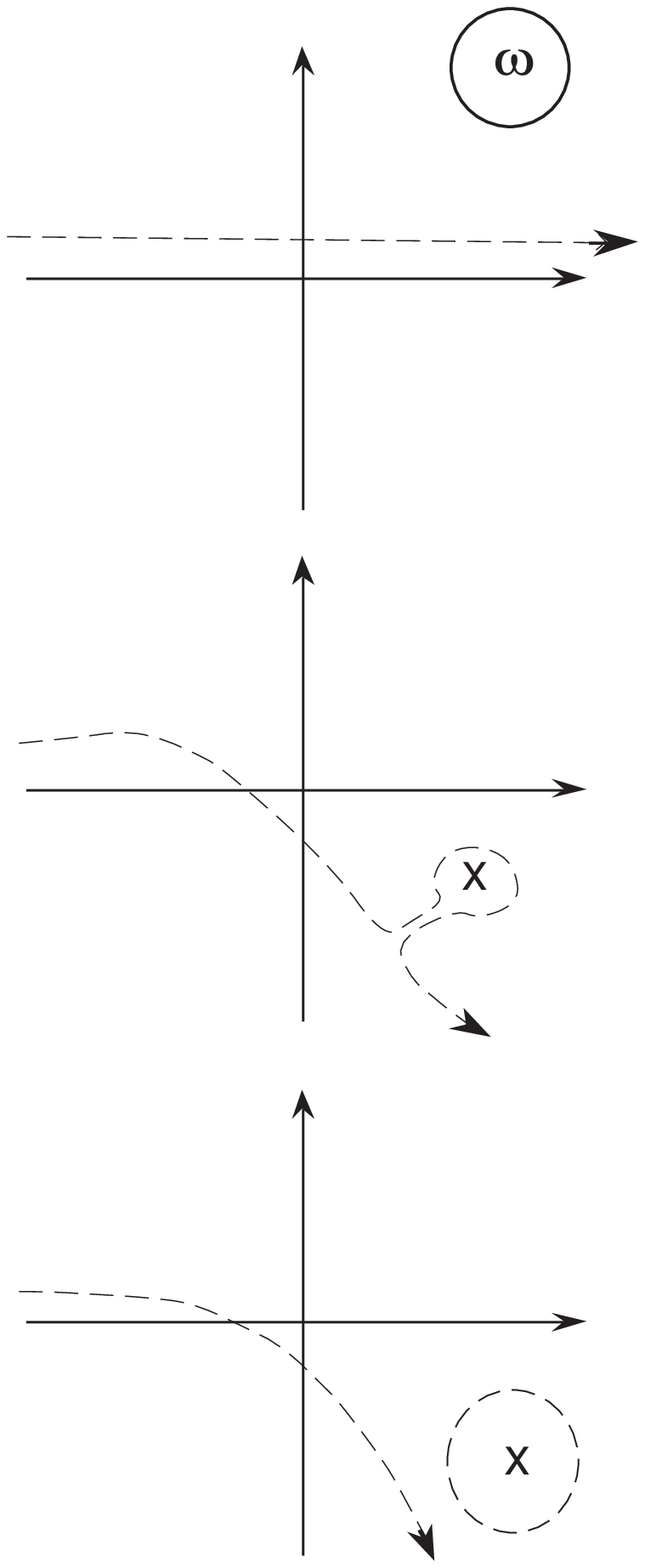}}
\caption[f1]{In the complex $\omega $-plane, the Laplace-Fourier transform is
inverted by integration along the line $[-\infty +ia,+\infty +ia].$ The analytic
continuation can be effected by moving the contour into the lower
half-plane. When a pole is encountered, the contour must circle it, but can
continue down leaving the pole encircled by an isolated contour.}
\end{figure}

In the region where $D(\omega)$ is an analytic function of $\omega$, we
can achieve the analytic continuation of the self-energy by defining the
new function
\begin{equation}
\Sigma ^{+}(\omega )=\int d\epsilon \left( \frac{D(\epsilon )}{\omega
-\epsilon }\right) -2\pi iD(\omega ). \label{a24} \end{equation}
Using (\ref{a20}) one can easily verify that the function $\Sigma
^{+}(\omega )$ has no singularity when the real axis is crossed from above.
When this process works, it is a simple way to look for singularities on
the second sheet of the lower half plane.

When the result (\ref{a24}) is inserted into the equation (\ref{a18}) we
find the relation
\begin{equation}
\omega ^{^{\prime }}-\omega_0-\int d\epsilon \,D(\epsilon )\left( \frac{
\omega ^{^{\prime }}-\epsilon }{\left( \omega ^{^{\prime }}-\epsilon
\right) ^2+\omega ^{^{\prime \prime }2}}\right) +2\pi \frac{\partial
D(\omega ^{^{\prime }})}{\partial \omega }\omega ^{^{\prime \prime }}=0,
\label{a25} \end{equation}
where we have anticipated that the imaginary part is small. In addition we
have
\begin{equation}
\omega ^{^{\prime \prime }}=2\pi iD(\omega ^{^{\prime }})-\int d\epsilon
\,D(\epsilon )\left( \frac{\omega ^{^{\prime \prime }}}{\left( \omega
^{^{\prime }}-\epsilon \right) ^2+\omega ^{^{\prime \prime }2}}\right) .
\label{a26}
\end{equation}
The values of the oscillational frequency $\omega ^{^{\prime }}$ and the
damping $\omega ^{^{\prime \prime }}$ are determined by the coupled
equations (\ref{a25}) and (\ref{a26}). The oscillational part determined
from (\ref{a25}) is discussed in detail by Cohen-Tannoudji~\cite{rrr1}. In
the limit $\omega ^{^{\prime \prime }}\rightarrow 0$ the results simplify.
When the damping is small, we obtain
\begin{equation}
\omega ^{^{\prime \prime }}\Rightarrow \pi D(\omega ), \label{a26a}
\end{equation}
which directly provides a consistency check. The Weisskopf-Wigner result
thus emerges in the weak coupling limit from our prescription for analytic
continuation.

It is instructive to realize that in the weak coupling limit, the
prescription we have used for analytic continuation is to replace $\Sigma
(\omega )$ just below the real axis by its value just above; i.e. we are
performing an analytic continuation using only the zeroth order term in a
Taylor expansion across the real axis. Thus the analytically continued
equation (\ref{a22}) becomes
\begin{equation}
\omega ^{\prime }-i\omega ^{\prime \prime }-\omega_0-\int d\epsilon
\;\frac{ D(\epsilon )}{\omega ^{\prime }+i\omega ^{\prime \prime
}-\epsilon }=0. \label{a27}
\end{equation}
For small values of $\omega ^{\prime \prime }$ this gives back all results
derived previously. It is to be noted that, inside the integral, the
precise value of $\omega ^{\prime \prime }$ does not affect the result, and
we can obtain the correct damping using the conventional infinitesimal
$i\eta $ prescription. Our discussion is in fact a derivation of this rule. 

An instructive way to consider the equation (\ref{a27}) is to regard the
imaginary part in the denominator as a small dissipative part of the
energies in the continuum
\begin{equation}
\epsilon -i\omega ^{\prime \prime }\Rightarrow \epsilon -i\epsilon
^{^{\prime }}. \label{a28}
\end{equation}
This provides an interesting physical interpretation of the results. Adding
any minute dissipative mechanism to the continuum we will find that the
isolated state will decay with the rate (\ref{a26a}) independently of the
details of the dissipation of the continuum. Thus any system coupled
perturbatively to a dissipative probability sink through a continuum will
decay with a rate determined by the coupling strength and not by the actual
dissipation of the continuum. In the perturbative limit, the probability
flow into the continuum will be the bottleneck. The dissipated continuum
acts as a proper reservoir unable to return the probability once it has
received it. This interpretation of the procedure directly provides the
right sign in the denominator of (\ref{a27}) to effect the proper analytic
continuation.

When we calculate the integral in Eq.~(\ref{a17}), we distort the contour
into the lower half plane until we encounter the pole defined by
(\ref{a25}) and separate a contour $c_0$ circling this. If no other
singularity were encountered, we could pull the contour to negative
imaginary infinity and find no corrections. In a physical system this is,
however, impossible because the spectral density $D(\omega )$ has
necessarily got the lower limit $\mu $ as explained above. The branch cut
starting here is pulled down to $-i\infty $ in the manner shown in Fig.~2.
This contributes an additional contour $c_1$ as shown in the figure. Thus
we find 
\begin{eqnarray}
A(t) & = & \frac{i}{2\pi}\left[ \int_{c_1} dt\,e^{-i\omega t}\;\left(
\frac{1}{\omega -\omega_0-\Sigma (\omega )}\right)
+\int_{c_0} dt\,e^{-i\omega t}\;\left( \frac{1}{\omega -\omega
^{^{\prime }}+i\omega ^{^{\prime \prime }}}\right) \right] \nonumber\\
& = & I_1(t)+\exp \left( -i\omega ^{^{\prime }}t-\omega^{^{\prime \prime
}}t\right) .\label{a29}
\end{eqnarray}
The second term is as expected in a perturbative calculation of the
Weisskopf-Wigner type, and we have taken the residue $Z$ of (\ref{c4}) to
be unity. The decay rate of the probability is given by the well known
expression
\begin{equation}
\gamma =2\omega ^{^{\prime \prime }}=2\pi D(\omega ). \label{a30}
\end{equation}
However, the corrections from the cut are given as 
\begin{equation}
I_1(t)=\frac{e^{-i\mu t}}{2\pi }\int\limits_0^\infty e^{-\xi t}\left(
\frac{ \Sigma_I(\mu -i\xi +\eta )-\Sigma_{II}(\mu -i\xi -\eta )}{\left(
\omega -\omega_0-\Sigma_I\right) \left( \omega -\omega_0
-\Sigma_{II}\right) } \right) d\xi , \label{a31}
\end{equation}
where the infinitesimal parameter $\eta $ ensures that the self energy is
evaluated on the two sides of the cut. From this form we can see that in
the long time limit $t\rightarrow \infty $, only the value $\xi \approx 0$
contributes. This would give no contribution without the singularity of $
\Sigma $ across the cut. Thus we can set $\xi =0$ everywhere except in the
integral
\begin{eqnarray}
\Sigma_I(\mu -i\xi +\eta )-\Sigma_{II}(\mu -i\xi -\eta ) & = &
\int_{\mu}^{\infty} d\epsilon \;D(\epsilon )\left[ \frac{1}{\mu -i\xi
+\eta -\epsilon }-\frac{1}{\mu -i\xi -\eta -\epsilon }\right] \nonumber\\
& = & \int\limits_0^\infty d\nu \;D(\mu +i\nu )\left[ \frac{1}{\xi +\nu
+i\eta }-\frac{1}{\xi +\nu -i\eta }\right] \label{a32}\\
& = & -2\pi iD(\mu +i\xi ).\nonumber
\end{eqnarray}
Because $\omega =\mu $ is a threshold, we expect that we have
\begin{equation}
D(\mu +i\xi )=\beta \;\left( i\xi \right) ^\alpha +... \label{a33}
\end{equation}
for small values of $\xi .$ Inserting this into the equation (\ref{a31}),
we find the leading term
\begin{equation}
I_1(t)=\frac{\beta \,e^{-i\mu t}\;i^{\alpha +3}}{\left( \mu -\omega_0
-\Sigma (\mu )\right) ^2}\Gamma (\alpha +1)\left( \frac 1{t^{\alpha
+1}}\right) . \label{a34}
\end{equation}
This type of asymptotic result was given for scattering theory already in
Ref.~\cite{r7}. When the exponential in (\ref{a29}) has decayed, the power
dependence (\ref{a34}) will dominate for long times.

\begin{figure}
\vspace*{-0.5cm}
\centerline{\psfig{width=3.4in,file=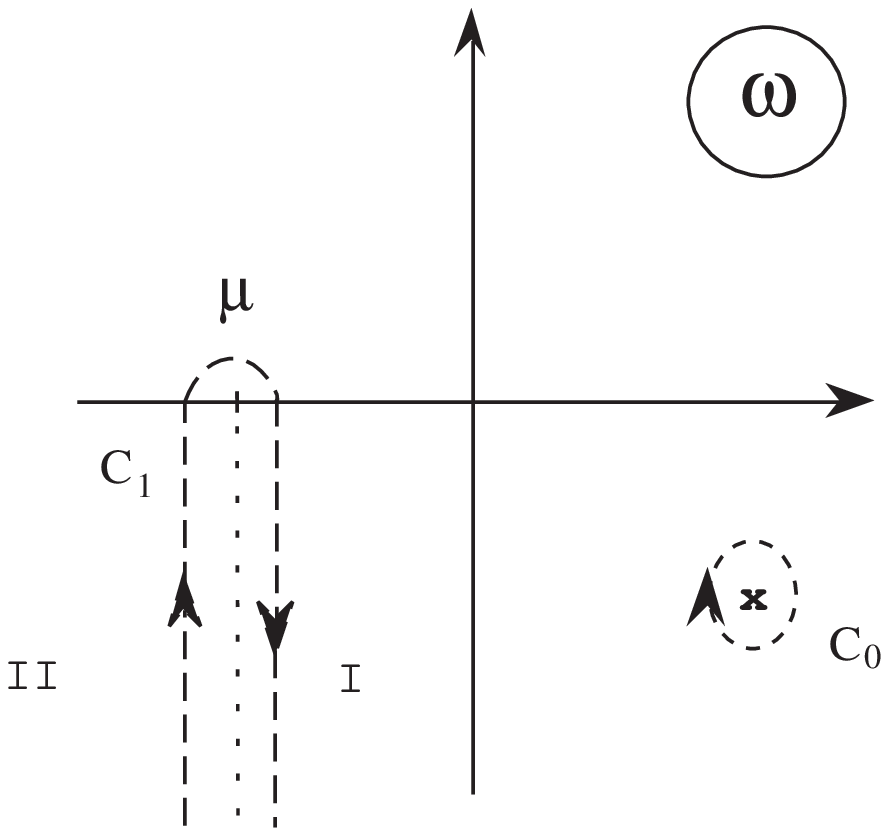}}
\caption[f2]{Because physical spectra have a lower limit $\mu $, there must be a
branch cut starting at this value. When isolated poles are encircled by closed
contours, here $c_0$, the branch cut can be drawn from $\mu $ to $-\infty$,
thus contributing another contour $c_1$.}
\end{figure}

The treatment given above does, of course, assume that there are no
additional poles encountered. If these are well separated, they contribute
simply additional exponential terms, but for poles situated close to each
other interferences occur; for a discussion of these cf. the discussion by
Mower~\cite{r8}.

\section{Simple models}

\subsection{A pole in the spectral density} 

Many systems display a moderately sharp spectral feature in its density of
states. Hence one may often approximate this with a Lorentzian shape, which
simplifies the treatment considerably. Such assumptions are often utilized
to describe scattering from condensed matter~\cite{rrr2}, but it is useful
in theoretical discussions too. Recently Garraway has utilized it to
describe non-Markovian effects in terms of pseudomodes~\cite{r14}. 

We assume that the spectral function has the simple shape \begin{equation}
D(\epsilon )=\frac{A^2}{\left( \epsilon -a\right) ^2+b^2}. \label{a35}
\end{equation}
This satisfies the positivity requirement, but it lacks the lower limit
necessary in physical systems. Thus no power law terms are expected. 

With the expression (\ref{a35}), the analytic continuation to the second
sheet is trivial. Inserted into Eqs. (\ref{a16}) and (\ref{a15a}) it gives
the results
\begin{equation}
\begin{array}{lll}
\Sigma (\omega ) & = & \frac{\textstyle\pi A^2}{\textstyle b}\left(
\frac{\textstyle 1}{\textstyle\omega -a+ib}\right)
\\
& & \\
\langle 0\mid G(\omega )\mid 0\rangle & = & \frac{\textstyle\omega -a+ib}
{\textstyle (
\omega -a+ib)( \omega -\omega_0) -\frac{\textstyle\pi A^2}{\textstyle b}}.
\end{array}
\label{a36}
\end{equation}
This form is found to satisfy the correct initial condition
\begin{equation}
\lim_{\omega \rightarrow \infty }\omega \langle 0\mid G(\omega )\mid
0\rangle \,\psi (0)=\psi (0). \label{a37} \end{equation}
The general solution of the time evolution becomes \begin{equation}
A(t)=\left( R_1\,e^{i\Omega_{+}t}+R_2\,e^{i\Omega_{-}t}\right) ,
\label{a38}
\end{equation}
where $\Omega_{\pm }$ are the two roots of the denominator in
Eq.~(\ref{a36}); from the form of the equation it follows that the sum of
the residues is unity: $R_1+R_2=1$.

Inserting the ansatz (\ref{a21}) into the denominator of Eq.~(\ref{a36}) we
find that the imaginary part follows as the solution of the equation
\begin{equation}
\omega ^{^{\prime \prime }}=\left( \frac{\pi A^2}{b}\right) \frac{\left(
b-\omega ^{^{\prime \prime }}\right) }{\left( a-\omega ^{^{\prime }}\right)
^2+\left( b-\omega ^{^{\prime \prime }}\right) ^2}. \label{a39}
\end{equation}
Because the right hand side of this equation is positive for $\omega
^{^{\prime \prime }}<b$ there must be a solution for $0<\omega ^{^{\prime
\prime }}<b$, irrespective of the value of $\omega ^{^{\prime }}.$ This
proves that both roots $\Omega_{\pm }$ have negative imaginary parts and
the solution $A(t)$ decays to zero as we expect. Thus the pole
approximation is consistent and only fails to reproduce the long time
behaviour deriving from the threshold in the spectral density. In a similar
manner the appearance of several poles can be discussed. Of special
interest is the case when a pole with negative weight is encountered. This
case is discussed in detail by Garraway~\cite{r14a}.

\subsection{The branch cut model}

In order to illustrate the concepts introduced in Sec.~3 we introduce a
model where the Weisskopf-Wigner result emerges as an exact consequence. It
is then possible to follow how the analytic continuation to the physical
sheet works in detail, and the correction terms can be evaluated
explicitly. 

The model is based on the following spectral density 
\begin{equation}
D(\epsilon )=\left\{
\begin{array}{cc}
A^2; & \mid \epsilon \mid <L \\  & \\
0; & \mid \epsilon \mid >L.
\end{array}\right.
\label{a40}
\end{equation}
It is taken to be an essential feature of the model that $L$ is taken to be
large (infinite) in the final expressions. For finite $L,$ it is trivial to
compute the self-energy from (\ref{a16}) to be \begin{equation}
\Sigma (\omega )=A^2\log \left( \frac{-\left( L+\omega \right) }{\left(
L-\omega \right) }\right) . \label{a41}
\end{equation}
In this model we can explicitly see that the sign of the imaginary part
clearly depends on the choice of branch of the function involved. We use
the discussion in Sec.~3 to introduce the analytic continuation (\ref{a24})
\begin{equation}
\Sigma ^{+}(\omega )=A^2\left[ \log \left( \frac{-\left( 1+\frac \omega
L\right) }{\left( 1-\frac \omega L\right) }\right) -2\pi i\right] .
\label{a42}
\end{equation}
If we now introduce the ansatz
\begin{equation}
\omega =\omega ^{^{\prime }}-i\omega ^{^{\prime \prime }}. \label{a43}
\end{equation}
we find that
\begin{equation}
\log \left( \frac{-\left( 1+\frac \omega L\right) }{\left( 1-\frac \omega
L\right) }\right) =\log \left( \frac{R_{-}}{R_{+}}\right) +i\left( 
\varphi_{-}-\varphi_{+}\right) , \label{a44}
\end{equation}
where we have
\begin{equation}
\frac{R_{-}}{R_{+}}=\sqrt{\frac{\left( 1+\frac{\omega ^{^{\prime
}}}L\right) ^2+\left( \frac{\omega ^{^{\prime \prime }}}L\right) ^2}{\left(
1-\frac{ \omega ^{^{\prime }}}L\right) ^2+\left( \frac{\omega ^{^{\prime
\prime }}} L\right) ^2}} \label{a45}
\end{equation}
and
\begin{equation}
\varphi_{\mp }=\arctan \left( \frac{\omega ^{^{\prime \prime }}}{\left(
L\pm \omega ^{^{\prime }}\right) }\right) . \label{a46} \end{equation}
These relations are illustrated in Fig.~3. It is now easy to see that
independently of whether we choose the cut of the logarithm function from
the origin to $+\infty $ or to $-\infty ,$ the limit becomes
\begin{equation}
\lim_{L\rightarrow \infty }\Sigma ^{+}(\omega )=-i\,\pi A^2, \label{a47}
\end{equation}
giving exactly the decay constant
\begin{equation}
\gamma =2\pi A^2; \label{a48}
\end{equation}
cf. Eq.~(\ref{a30}). The frequency shift vanishes in this limit. Our
treatment is not mathematically rigorous, and the contributions from the
branch cut are not evaluated. They disappear when the lower spectral limit
$ -L$ goes to infinity. It is, however, possible to evaluate the
correction terms to order $L^{-1}$ and verify the over-all validity of the
picture we have developed. The model can also be extended to treat an
asymmetric spectral weight function defined on the interval
$[L_{-},L_{+}].$ Here the limit appears slightly differently and
modifications of the results above can be found.

\begin{figure}[tbh]
\hspace*{-2cm}\psfig{width=6.4in,file=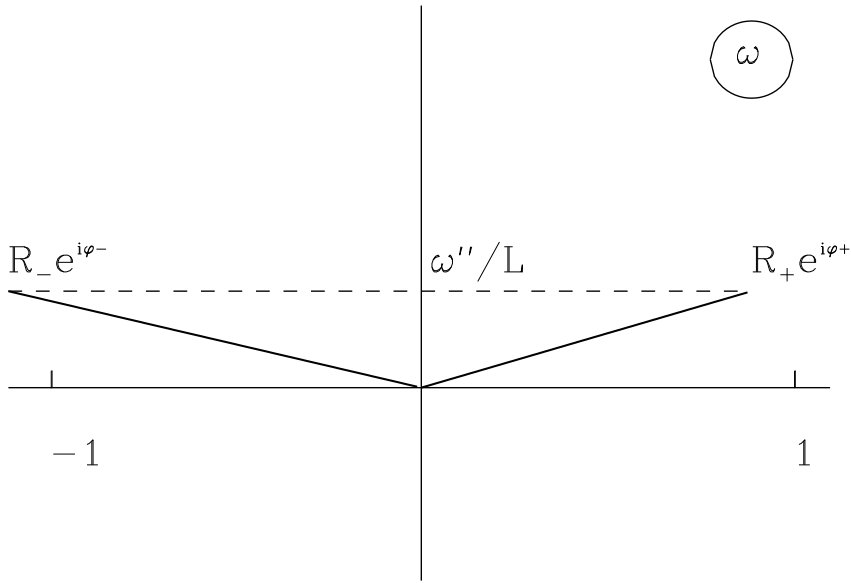}

\vspace*{-2.5cm}
\caption[f3]{The branch-cut model gives the self energy $\Sigma$ in terms of 
two vectors $R_{\pm }\exp \left( i\varphi_{\pm }\right)$ in the complex plane
as shown.}
\end{figure}

With the result (\ref{a48}) we find an exact exponential disappearances of
the amplitude of the initial state in Eq.~(\ref{a17}). It is, however,
instructive to evaluate the state in the continuum part too. The
probability of emergence of this state cannot grow faster than the isolated
state disappears, and consequently we expect it to need a time of the order
$ \gamma ^{-1}$ to appear. From Eq.~(\ref{a15b}) we find for the continuum
wave function
\begin{eqnarray}
\Psi_{cont}(x,t) & = & \int \int d\epsilon \,\phi_\epsilon (x)\langle
\epsilon \mid G(\omega )\mid 0\rangle \;e^{-i\omega t}\frac{id\omega
}{2\pi } \nonumber\\
& = & \int d\epsilon \,\phi_\epsilon (x)\left( \frac{V_\epsilon
}{\epsilon -\omega_0+i\frac \gamma 2}\right) \left( e^{-i\epsilon
\,t}-e^{-i\omega_0t}e^{-\gamma t/2}\right) \label{a49} \\
& = & \Phi (x,t)-e^{-i\omega_0t}e^{-\gamma t/2}\Phi (x,0). \nonumber
\end{eqnarray}
Here $\phi_\epsilon (x)$ is the continuum eigenfunction corresponding to
the eigenvalue $\epsilon $, and $\Phi (x,t)$ is the state emerging from the
process on the corresponding Hilbert space. This has got the spectral
distribution
\begin{equation}
P(\epsilon )=\frac{\gamma /2\pi }{\left( \epsilon -\omega_0\right)
^2+\left( \frac \gamma 2\right) ^2} \label{a50} \end{equation}
as known from the Weisskopf-Wigner calculation. In addition, we can see how
the continuum state grows from an initial zero value to the emerging wave
state $\Phi (x,t),$ which will travel according to the dynamics in the
continuum as determined by the spectral distribution (\ref{a50}). This
feature has usually not been discussed in the ordinary decay problems
treated in the literature.

In the present model the analytic behaviour can be seen directly. This is
useful if we want to gain understanding of the features playing the central
role in the general treatment. We can also calculate the emergence of the
outgoing wave packet explicitly (\ref{a49}). In order to see the shaping of
this wave packet we consider one more model where the features discussed
can be followed numerically.

\subsection{A wave packet model}

\begin{figure}[tbh]
\hspace*{-2cm}\psfig{width=6.4in,file=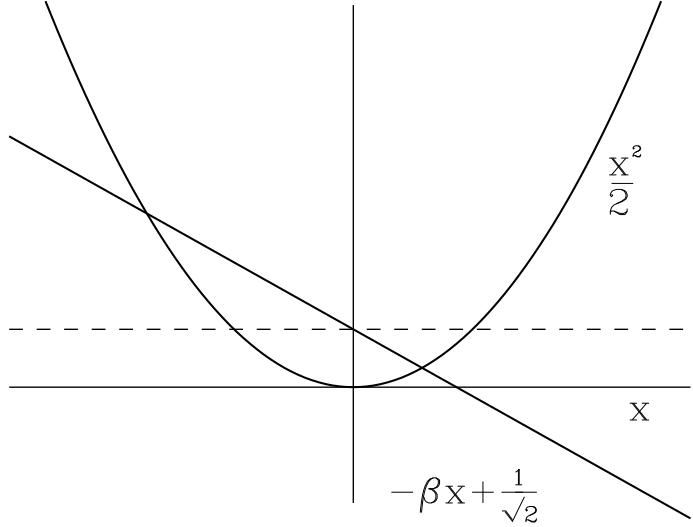}

\vspace*{-2.5cm}
\caption[f4]{The molecular energy levels have one harmonic potential and one
straight slope. The parameters are chosen such that the energies of the two
levels coincide at $x=0$.}
\end{figure}

As a model we choose a coupled pair of energy levels with one harmonic
potential and one potential slope providing a continuum. The Schr\"odinger
equation is
\begin{equation}
  i\frac \partial {\partial t}\left[
  \begin{array}{c}
    \psi_1 \\ \psi_2
  \end{array} \right] =\left[
  \begin{array}{cc}
     \left( -\frac{\partial ^2}{\partial x^2}+\frac 12x^2\right) & V \\ V &
     \left( -\frac{\partial ^2}{\partial x^2}-\beta x+\frac 1{\sqrt{2}}\right)
   \end{array}\right] \left[
  \begin{array}{c}
    \psi_1 \\ \psi_2
  \end{array}\right] . \label{a51}
\end{equation}
This corresponds to a scaled molecular problem with the parameters
\begin{equation}
  m=\frac 12; \qquad \omega =\sqrt{2}. \label{a52}
\end{equation}
Because the zero point energy is given by \begin{equation}
\frac 12\omega =\frac 1{\sqrt{2}}, \label{a53} \end{equation}
the ground state energy of the harmonic oscillator coincides with the
continuum energy at the origin; see Fig.~4. To the extent we can neglect the
influence of the first excited harmonic oscillator state, we can consider
this model as a realization of a situation where the initial oscillator
ground state 
\begin{equation}
\psi_1(x,t=0)=(\pi \sqrt{2})^{-1/4}\exp \left(
-\frac{x^2}{2\sqrt{2}}\right) \label{a54}
\end{equation}
is embedded in the continuous spectrum of the slope; the coupling is given
by $V.$ When this is in the perturbative regime \begin{equation}
\frac{V^2}\beta \ll \omega \label{a55}
\end{equation}
we expect the Weisskopf-Wigner treatment to hold and the initial state to
decay nearly exponentially. Setting
\begin{equation}
   V=0.5; \qquad \beta =3.0 \label{a56}
\end{equation}
we satisfy (\ref{a55}) and obtain the decay of the initial state
\begin{equation}
P_1(t)=\int \mid \psi_1(x,t)\mid ^2dx \label{a57} \end{equation}
shown in Fig.~5. As seen, there is a clear range of times over which the
initial state decays exponentially as expected. By looking at the state
emerging on the second energy level $\psi_2(x,t)$ shown in Fig.~6 we can
see that the emerging state like (\ref{a49}) indeed produces a wave packet
which travels down the potential slope while it spreads according to the
requirements of quantum mechanics. In the present model, however, the
initial state is not depleted totally by the exponential decay. This
derives from the fact that the crossing potential surfaces form quasibound
states in the adiabatic representation, and the population trapped in these
states can then ooze out only over times much longer than those
characterizing the exponential decay.

The model calculation presented in this section shows that our system (\ref
{a51}) can serve as an illustration of Weisskopf-Wigner decay, and it
proves that such decay can serve as a wave packet preparation device.
However, we have found that even this simple system works only when the
parameters are chosen right, and it contains features not expected from the
simpler models discussed above. We consider these aspects in a forthcoming
publication, where we shall present the general behaviour of the harmonic
state resonantly coupled to a sloped potential. Here we have only presented
a single case as an illustration of our general theoretical considerations. 

\section{Discussion}\setcounter{equation}{0}

\begin{figure}[tbh]
\vspace*{-0.5cm}
\centerline{\psfig{width=4.8in,file=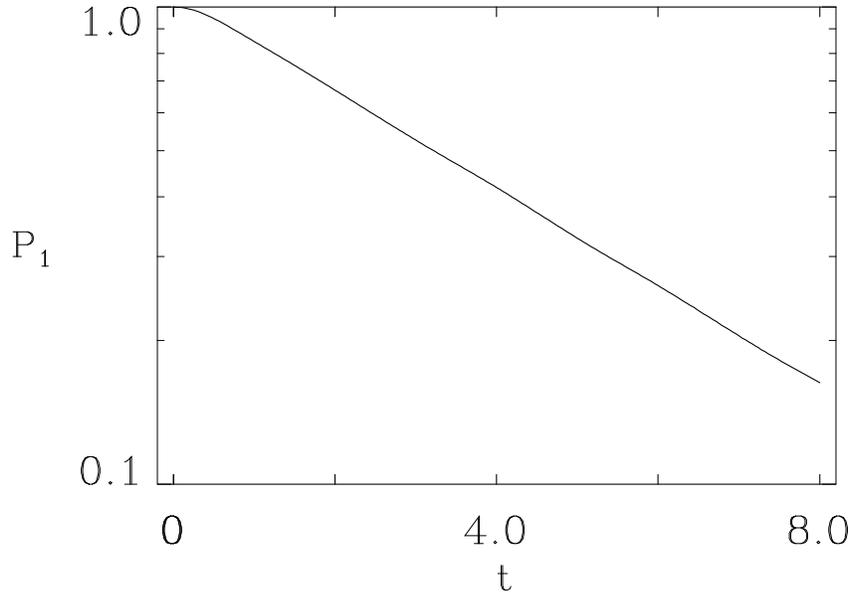}}
\caption[f5]{The system is prepared on the ground state of the harmonic 
potential,
and the probability that it remains there is plotted as a function of time. The
probability leaves the level in an exponential manner as this
semilogarithmic plot shows.}
\end{figure}

In this paper we have reviewed the scattering theory approach to
partitioning of the Hamiltonian time evolution in a quantum system. We are
especially interested in the situation where one (or more) discrete states
leak into a continuum, i.e., decay of quasistationary states. By choosing a
Laplace-Fourier transform in time, we concentrate on solving an initial
value problem and following the system as it irreversibly transfers its
probability into the continuum. This is regarded as a wave packet
preparation procedure.

The emphasis on time evolution has been motivated by the recent
experimental progress in pulsed laser technology. It is now possible to
excite a single state selectively, control its coupling to other states,
including continuum ones, and follow the ensuing time evolution of the
quantum states. Such work has experienced tremendous progress in molecular
investigations recently, but also time resolved spectroscopy of
semiconductor structures is possible. 

Starting from a giving initial time, the future evolution of quantum
systems is determined by the analytic properties of the propagator
functions in the complex frequency plane. In order to obtain the correct
physical behaviour these have to be continued analytically to a second
Riemann sheet in the lower half plane. Any pole encountered will give rise
to a resonance but the physical behaviour must also receive contributions
from unavoidable branch cuts.

\begin{figure}[tbh]
\vspace*{-0.5cm}
\centerline{\psfig{width=4.5in,file=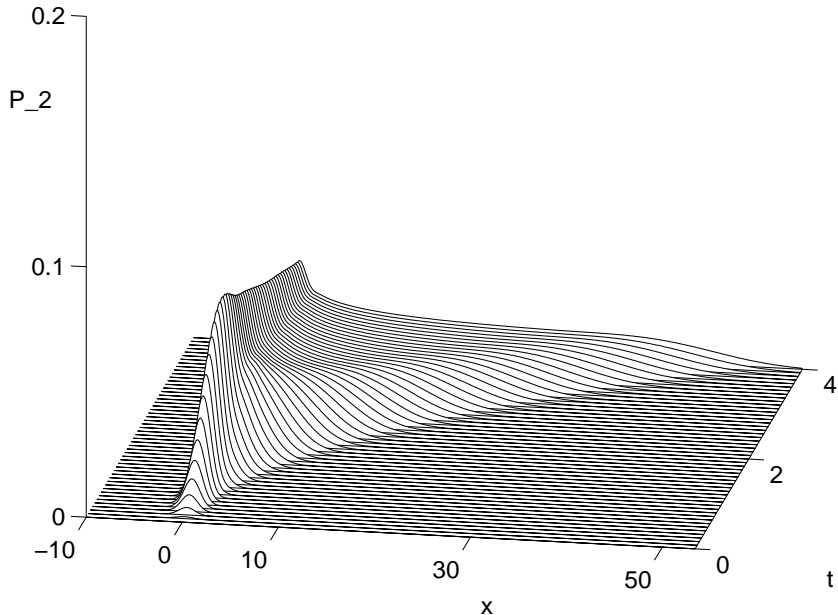}}
\caption[f6]{The probability $|\psi_2(x,t)|^2$ leaving the bound oscillator
state appears on the sloping continuum potential. Here we can see it appear
and fall down the slope in the form of a spreading wave packet. Part of the
probability gets captured near the position of the original bound state
because of trapping in the potential well formed by the adiabatic levels.
This contains a component on the continuous manifold, which explains the
trapped portion. Even this must eventually leave the trap but clearly over
a much longer time scale.}
\end{figure}

The necessary analytic continuation introduces simple relations between the
self-energy function in the upper half-plane and that in the lower. In the
weak coupling limit, the Weisskopf-Wigner case, the prescription for
continuation is found to be equivalent with adding a small dissipative
mechanism to the states in the continuum. The causality requirement on the
sign of this dissipation automatically affects the necessary analytic
continuation. Physically we can understand this as follows: When
probability leaks into the continuum through the weak coupling link, it is
dissipated by all modes and oozes into the unobserved degrees of freedom
providing the continuum damping. In scattering theory this mechanism is
simply the outgoing boundary condition at infinity; all scattered modes are
irreversibly lost at the edges of our universe. The bottle neck is provided
by the weak link, and the rate of loss of the initial state is fixed by
this. The dissipation in the continuum is only present to prevent any
return of probability. Hence its details are inessential to the decay, its
presence however is necessary.

We have discussed the analytic behaviour of the propagators in detail, and
illustrate the properties by simple model calculations. Most of the
treatment follows the tradition in this field and carries out the argument
in the energy representation. However, to prove the existence of the
phenomena discussed, we display the behaviour of a model describing laser
coupling of a bound molecular state into a dissociating continuum. In
addition to showing the expected exponential decay, the model system proves
that the state prepared on the continuum is in the form of an outgoing wave
packet. The model, however, also displays features not simply describable
in the model calculations performed in this paper; we will return to a
detailed exploration of the decay characteristics in a forthcoming
publication. 

The purpose of this paper is to draw attention to the necessity to
investigate the time evolution in coupled quantum systems. This provides a
broad range of interesting quantum models, which may also shed light on the
phenomena occurring in dynamical processes initiated and explored by pulsed
lasers.

\appendix

\section*{Appendix}

\def\theequation{A.\arabic{equation}}
\setcounter{equation}{0}

The algebraic manipulations we need in this paper have been provided by
scattering theory long ago~\cite{r7}. Here we summarize the main results
only.

The perturbation series for the propagator is generated by the relation
\begin{equation}
G=G_0+G_0VG. \label{x1}
\end{equation}
Introducing the operator $T$ by the relation \begin{equation}
VG=TG_0, \label{x2}
\end{equation}
we find the equation
\begin{equation}
G=G_0+G_0TG_0. \label{x2a}
\end{equation}
The operator $T$ is obtained from the equation 
\begin{equation}
T=V+TG_0V. \label{x3}
\end{equation}
The disconnected operator $\Gamma $ is defined by the relation
\begin{equation}
\Gamma =V+VG_0^Q\Gamma , \label{x4}
\end{equation}
where the projector $Q=1-P$ has been introduced. The missing singularities
in the intermediate states can be reintroduced to get the $T$ operator
\begin{equation}
T=\Gamma +TG_0^P\Gamma . \label{x5}
\end{equation}

When we remember the projectors commute with $H,$ we can easily solve
\begin{equation}
T^P=\Gamma ^P\left( \frac 1{1-G_0^P\Gamma ^P}\right) \label{x6}
\end{equation}
and
\begin{equation}
\begin{array}{ccc}
G^P & = & G_0^P\left( 1+T^PG_0^P\right) \\ & & \\
& = & \left( \frac 1{\left( G_0^p\right) ^{-1}-\Gamma ^P}\right) .
\end{array}
\label{x7}
\end{equation}
In the same way we can calculate $G^{QP}$ from Eq.~(\ref{x1}) and $G^Q$ from
Eq.~(\ref{x2a}). The results are given in Eqs.~(\ref{a12b}) and (\ref{a12c}). 

These results are all exact, and they can be used to solve for the
propagators of any Hamiltonian split into $H_0$ and $V$ and partitioned by
projectors $P$ and $Q$ which separate the Hilbert space into subspaces not
coupled by $V$.

\setlength{\baselineskip}{4mm}
\renewcommand{\baselinestretch}{0.8} 
\setlength{\itemsep}{0.0mm}
\setlength{\parsep}{2.0mm}

\end{document}